# Geometrical Magnetic Frustration in Rare Earth Chalcogenide Spinels

G.C. Lau[1], R.S. Freitas[2], B.G. Ueland[2], P. Schiffer[2], and R.J. Cava[1]

[1]Department of Chemistry, Princeton University, Princeton, NJ 08544

[2]Department of Physics and Materials Research Institute, Pennsylvania State University, University Park, PA 16802

**Abstract**

We have characterized the magnetic and structural properties of the $CdLn_2Se_4$ (Ln = Dy, Ho), and $CdLn_2S_4$ (Ln = Ho, Er, Tm, Yb) spinels. We observe all compounds to be normal spinels, possessing a geometrically frustrated sublattice of lanthanide atoms with no observable structural disorder. Fits to the high temperature magnetic susceptibilities indicate these materials to have effective antiferromagnetic interactions, with Curie-Weiss temperatures $\Theta_W \sim -10$ K, except $CdYb_2S_4$ for which $\Theta_W \sim -40$ K. The absence of magnetic long range order or glassiness above T = 1.8 K strongly suggests that these materials are a new venue in which to study the effects of strong geometrical frustration, potentially as rich in new physical phenomena as that of the pyrochlore oxides.

The geometry of the lanthanide ion sublattice in the pyrochlore structure is an array of corner sharing tetrahedra based on kagomé planes that connect via triangular layers. This arrangement of spins can show geometric frustration, such that no long range magnetic ordering occurs, even at low temperatures.[1,2] The magnetic behavior of the $Ln_2Ti_2O_7$ pyrochlores is widely studied, and, despite weak coupling between structural and magnetic degrees of freedom, the rare earth pyrochlores exhibit a particularly rich range of magnetic behavior due to a balance between dipole interactions, single ion anisotropy, and exchange couplings.[3,4,5,6,7,8,9] Many of these materials possess strong crystal fields causing highly anisotropic spin states. In $Dy_2Ti_2O_7$ and $Ho_2Ti_2O_7$ the spins' directionality and effective ferromagnetic interactions produce Ising-like spins leading to spin ice characteristics, where the orientations of the spins are analogous to the H positions in ice.[10,11,12,13,14,15,16,17,18,19,20,21,22] Exotic low temperature behavior has also been seen in $Gd_2Ti_2O_7$ [23], $Yb_2Ti_2O_7$ [24], and $Tb_2Ti_2O_7$ [7,8,9,25,26] among others in this class of materials.

The octahedral cation sublattice in spinels is identical in geometry to the lanthanide cation sublattice in pyrochlores. Transition metal spinel compounds such as $ZnCr_2O_4$ [27] and $CdFe_2O_4$ [28] are also widely studied in the context of geometric frustration. However, in contrast with the lanthanide pyrochlores, transition metal spinels possess strong couplings between the spin and lattice degrees of freedom, often leading to rich phenomena where magnetic ordering and structural transitions couple.[29,30] Here we report magnetic and structural characterizations of $CdLn_2Se_4$ (Ln = Dy, Ho) and $CdLn_2S_4$ (Ln = Ho, Er, Tm, Yb) spinels. Due to the presence of nonmagnetic atoms on the A sites and an absence of disorder in the distribution of the ions, these compounds have magnetic sublattices analogous to the pyrochlores. Previous work exists on many of these compounds,[31-43] and the findings here expand on what has been reported. In addition, the properties reported in the literature are in disagreement, and no prior investigations into their potential to show novel low temperature properties associated with the presence of strong geometrical frustration exist. The differences in lanthanide-anion coordination and site symmetry between the pyrochlores and spinels intuitively suggest that the magnetic behavior of the lanthanides in the spinels differs from that found in the equivalent pyrochlores. The potential exists, however, for as rich a manifold of novel magnetic behavior in the sub-Kelvin temperature regime.

We synthesized $CdLn_2S_4$ (Ln = Ho, Er, Tm, Yb) spinels by firing CdS and $Ln_2S_3$ ground together in an evacuated sealed quartz tube at 900°C for 3-5 days. CdS was synthesized from CdO (99.998%) heated at 300°C for 4 hrs under flowing argon bubbled through $CS_2$ (99.9%). Stoichiometric amounts of Ho (99.9%), Er (99.9%), Tm (99.9%), and Yb (99.9%) metals were each reacted separately with S (precipitated purified) in sealed quartz tubes at 800°C for 2 days to form $Ln_2S_3$. $CdLn_2Se_4$ (Ln = Dy, Ho) was made by reacting the elements (Dy - 99.9%) in evacuated sealed quartz tubes at 900°C for 2 weeks.

We characterized the sample structures through x-ray powder diffraction data using CuKα radiation and a diffracted beam monochromater. Structural refinements were made using the Bruker AXS software package TOPAS 2.1© operated with a Pseudo-Voight TCHZ fitting profile. Refined parameters include: zero corrections; sample displacement; scaling factors; cell dimensions; atomic positional coordinates; and thermal parameters ($B_{eq}$). We measured the d.c. magnetic susceptibility with a SQUID magnetometer (Quantum Design MPMS) on cooling over T = 290 - 2 K in an applied field of $H$ = 0.1 T. We also performed measurements of a.c. magnetic susceptibility using the ACMS option for a Quantum Design PPMS cryostat. This instrument also gives d.c. magnetization measurements in fields up to 9 tesla using an extraction method. Curie-Weiss fits to the d.c. susceptibility data performed over T = 80 – 270 K, with the exception of CdYb$_2$S$_4$, where fits were performed between T = 250 - 350 K. The applied field dependence of the magnetization was measured at selected temperatures for all materials up to a field of 7 Tesla.

Structure refinement shows all samples to be normal spinels with lattice parameters similar to those found previously[32, 35, 36]. Contrary to one previous report,[38] we found no evidence for mixing of the cations among the two metal sites: refinements of the occupancies of the A and B sites did not result in statistically better fits compared to the ordered normal spinel structure. Table 1 lists the atomic positions for CdEr$_2$S$_4$ as an example. Table 2 lists the measured lattice parameters (*a*), atomic positions *x* (where *x* = *y* = *z*) for S or Se, and selected bond distances and angles for all of the samples studied.

Despite an analogous magnetic sublattice geometry, the local bonding Ln-X polyhedra surrounding the lanthanides in the spinel are drastically different from those in the pyrochlore structure. Figure 1 compares the local bonding environment of Er in CdEr$_2$S$_4$, determined in our structure refinement, with the Er in the reported crystal structure of Er$_2$Ti$_2$O$_7$[44] (which is equivalent to that found in all of the frustrated pyrochlores with only slight differences in bond lengths). The sixfold sulfur coordination around Er in CdEr$_2$S$_4$ forms an almost ideal octahedron, with a slight trigonal distortion from the normal 90° bond angles. The Er-S bond distance is equivalent for all six sulfurs. This presents a stark contrast to the 8-fold oxygen coordination for Er in $Er_2Ti_2O_7$, where two very short Er-O bonds are found. Figure 1 shows that the short bonds in the pyrochlore point toward (and away from) the center of the lanthanide ion tetrahedra: this crystal field directs the Ising magnetic moments into the "in" and "out" configuration that is necessary for the display of spin ice behavior. The substantial difference in crystal fields between lanthanide spinels and pyrochlores suggests that their magnetic behavior will be quite different despite their analogous magnetic sublattice geometries. The longer Ln-Ln distance in the tetrahedral magnetic lattice in the spinel suggests weaker magnetic coupling than in the corresponding pyrochlore, but the values of $\theta_W$ are comparable and even sometimes larger, as described below.

Figure 2 and Figure 3 show the magnetic susceptibilities of $CdLn_2X_4$ spinels as inverse susceptibility versus temperature plots. The high temperature portion of each data set fits the Curie-Weiss law ($\frac{1}{\chi} = \frac{1}{C}(T - \theta_W)$, where $C$ is the Curie constant), yielding the $\theta_W$ intercepts and effective moments shown in Table 3. The experimental magnetic moments are consistent with the expected values for $Ln^{3+}$ ions. Our negative $\theta_W$ values are generally similar to those previously reported, among which there is considerable variation. It is notable that the $\theta_W$ values are negative for the Ho and Dy spinels (-7 to -8K) and much larger than those for the analogous spin ice pyrochlores.[21] This is possibly due to the influence of the substantially different crystal fields. Our Er spinel data is in contrast with a previous report, where the susceptibility of $CdEr_2S_4$ shows a turn over below 10 K.[32] $CdYb_2S_4$ has a much larger $\theta_W$ than the other spinels, and shows deviations from the Curie-Weiss fit below 150 K. This behavior results from the influence of low-lying crystal field levels, rather than strong exchange interactions.[40]

Given the antiferromagnetic interactions implied by the negative values of $\theta_W$, one might expect the onset of ordering at temperatures $\sim |\theta_W|$. The onset of such ordering is not seen in any of the samples studied, although, as shown in Figure 3, $CdTm_2S_4$ does display a decreasing slope in $M(T)$ at the lowest temperatures, possibly due to the dominance of Van Vleck susceptibility at those low temperatures.[32,41] There is no frequency dependence in the a.c. susceptibility or differences between the field-cooled and zero-field-cooled magnetization in any of the samples (data not shown). The absence of these signatures of glass type freezing excludes the possibility of a spin glass state above $T = 1.8$ K. The suppression of any sort of freezing or ordering down to temperatures well below $|\theta_W|$ confirms that the magnetism in these materials should be considered to be geometrically frustrated, as might have been guessed from their structure and the negative values of $\theta_W$.

We have also studied the field dependence of the magnetization, since there have been suggestions of interesting field-induced states in geometrically frustrated materials and demonstration of field induced states in single crystals of the spin ice materials.[15] Data taken on our materials at T = 2K up to a field of H = 9 T demonstrate that none display such behavior. They do not, however, approach the expected full saturated moment for the free ions. This indicates that the crystal field levels restrict the available spin states, as, for example, is the case for the frustrated spin ice compounds. In particular, $CdEr_2S_4$ shows a dramatic half magnetization plateau that is clearly developed at T = 2 K (figs. 4 and 5), reminiscent of what is seen in the spin ice compounds, $Dy_2Ti_2O_7$ and $Ho_2Ti_2O_7$.[15,21,22] A half magnetization plateau is also seen in $Er_2Ti_2O_7$,[21] which has been described as a realization of the frustrated <111> XY pyrochlore lattice antiferromagnet.[3] Our results suggest that the same may be true for $CdEr_2S_4$, but a detailed study of the crystal field levels is necessary to confirm this possibility.

In summary, we have investigated the possibility of geometrical frustration in the rare earth chalcogenide spinels. While our results do not indicate ordering of any sort, it is the absence of such behavior down to temperatures well below the scale expected from $\theta_W$ which indicates the importance of frustration to the magnetic behavior of these unique compounds. These materials present the opportunity for uncovering new physics as rich and complex as that observed in the frustrated oxide spinels and pyrochlores. Future studies to pursue such phenomena will probe at lower temperatures and in single crystal samples.

**Acknowledgement**

This research was supported by the National Science Foundation, under grant number DMR-035610. R.S.F. thanks the CNPq-Brazil for sponsorship. The authors gratefully acknowledge discussions with Shivaji Sondhi and Joseph Bhaseen.

Table 1.  Structural parameters for $CdEr_2S_4$ at room temperature;  Space group: Fd-3m (#227)  Lattice constant $a$ = 11.1178(4) Å

| Atom | position | $x$ | $y$ | $z$ | Occ | $B_{eq}$ |
|---|---|---|---|---|---|---|
| Cd | *8b* | 1/8 | 1/8 | 1/8 | 1 | 1.40(5) |
| Er | *16c* | 1/2 | 1/2 | 1/2 | 1 | 1.14(4) |
| S | *32e* | 0.2541(2) | 0.2541(2) | 0.2541(2) | 1 | 1.50(6) |

$\chi^2$ = 1.23;  $R_{wp}$ = 11.25;  $R_p$ = 8.61

Table 2.  Structural parameters for $CdLn_2X_4$ at room temperature;  Space group: Fd-3m (#227)

| Compound | Lattice constant, $a$ (Å) | S or Se position ($x$) | Cd-X distance (Å) | Ln-X distance (Å) | X-Ln-X bond angle (°) | $\chi^2$ | $R_{wp}$ (%) |
|---|---|---|---|---|---|---|---|
| $CdDy_2Se_4$ | 11.6467(9) | 0.2549(1) | 2.620(3) | 2.856(2) | 87.70(7) | 1.19 | 14.94 |
| $CdHo_2Se_4$ | 11.6273(5) | 0.2554(1) | 2.625(2) | 2.846(1) | 87.46(6) | 1.29 | 14.31 |
| $CdHo_2S_4$ | 11.1582(3) | 0.2544(2) | 2.501(3) | 2.741(2) | 87.92(8) | 1.25 | 13.29 |
| $CdEr_2S_4$ | 11.1178(4) | 0.2541(2) | 2.486(3) | 2.735(2) | 88.07(8) | 1.23 | 11.25 |
| $CdTm_2S_4$ | 11.0900(4) | 0.2555(2) | 2.507(3) | 2.713(2) | 87.38(8) | 1.19 | 12.24 |
| $CdYb_2S_4$ | 11.0562(3) | 0.2579(2) | 2.545(3) | 2.679(2) | 86.20(9) | 1.36 | 14.60 |

| Table 3.  Weiss constants and magnetic moments determined from the Curie-Weiss fits of high temperature portions of the magnetic susceptibilities. ||||
|---|---|---|---|
| Compound | $\theta_W$ (K) | p Exptl. | p Calc.$(g[J(J+1)]^{1/2})$ |
| $CdDy_2Se_4$ | -7.6 | 10.76 | 10.63 |
| $CdHo_2Se_4$ | -7.0 | 10.74 | 10.6 |
| $CdHo_2S_4$ | -7.6 | 10.60 | 10.6 |
| $CdEr_2S_4$ | -6.9 | 9.60 | 9.59 |
| $CdTm_2S_4$ | -11.8 | 7.58 | 7.57 |
| $CdYb_2S_4$ | -42.3 | 4.41 | 4.54 |

**Figures**

**Fig. 1** (Color on line) Upper portion: comparison of the Er-S coordination polyhedron in $CdEr_2S_4$ and the Er-O coordination polyhedron in $Er_2Ti_2O_7$[44]. Also shown are the lanthanide tetrahedra, illustrating the differences in the orientations of the Er-X crystal fields with respect to the magnetic lattice geometry. Lower portion: Observed and calculated powder X-ray diffraction pattern at ambient temperature for $CdEr_2S_4$. Lower line, difference between observed and calculated intensities.

**Fig. 2** (Color on line) Magnetic susceptibilities measured in an applied field of 1 kOe, plotted as inverse susceptibility vs. temperature, for $CdLn_2X_4$ spinels. Lines shown are the fits to the high temperature data. Inset: the data for $CdYb_2S_4$.

**Fig. 3** (Color on line) Magnetic susceptibility measured in an applied field of 1 kOe, plotted as inverse susceptibility vs. temperature for $CdTm_2S_4$ spinel. Line shown is the fit to the high temperature data. Inset: detail of linear susceptibility vs. temperature in the low temperature region for $CdTm_2S_4$.

**Fig. 4** (Color on line) Magnetization vs. applied field for two spinels, $CdTm_2S_4$ and $CdEr_2S_4$, measured to fields of 5T at three representative temperatures.

**Fig 5** (Color on line) Variation of the normalized magnetization with applied field for all spinels studied at 2K up to applied fields of 9T. The measured magnetization has been normalized by the value expected based on the effective number of Bohr magnetons determined in the high temperature susceptibility measurements.

**Figure 1.**

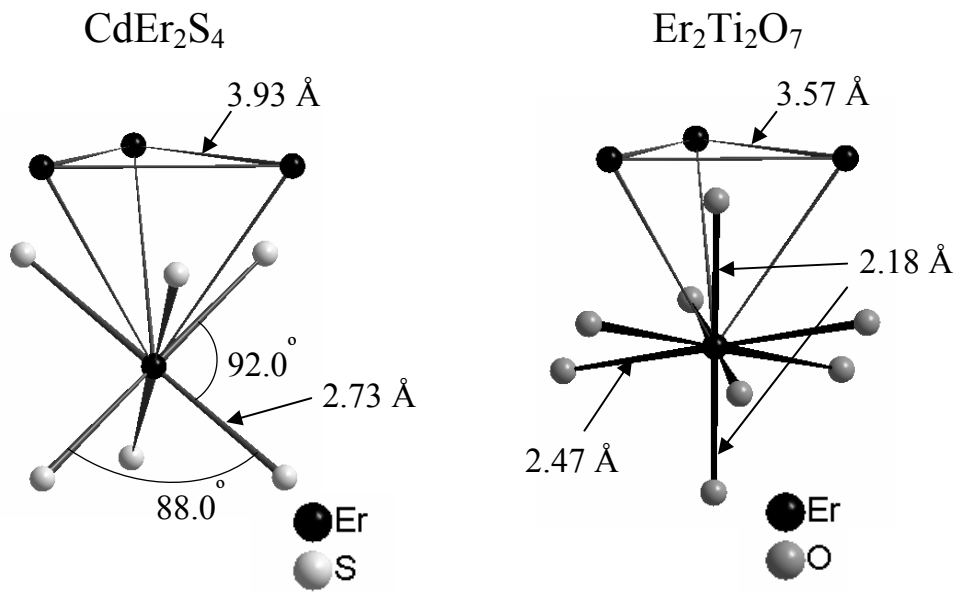

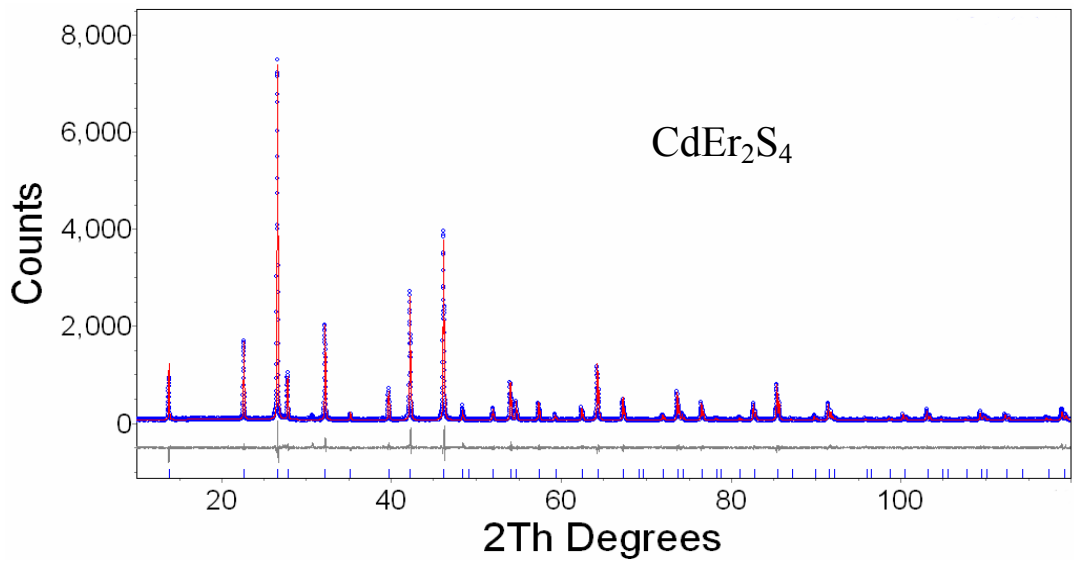

**Figure 2.**

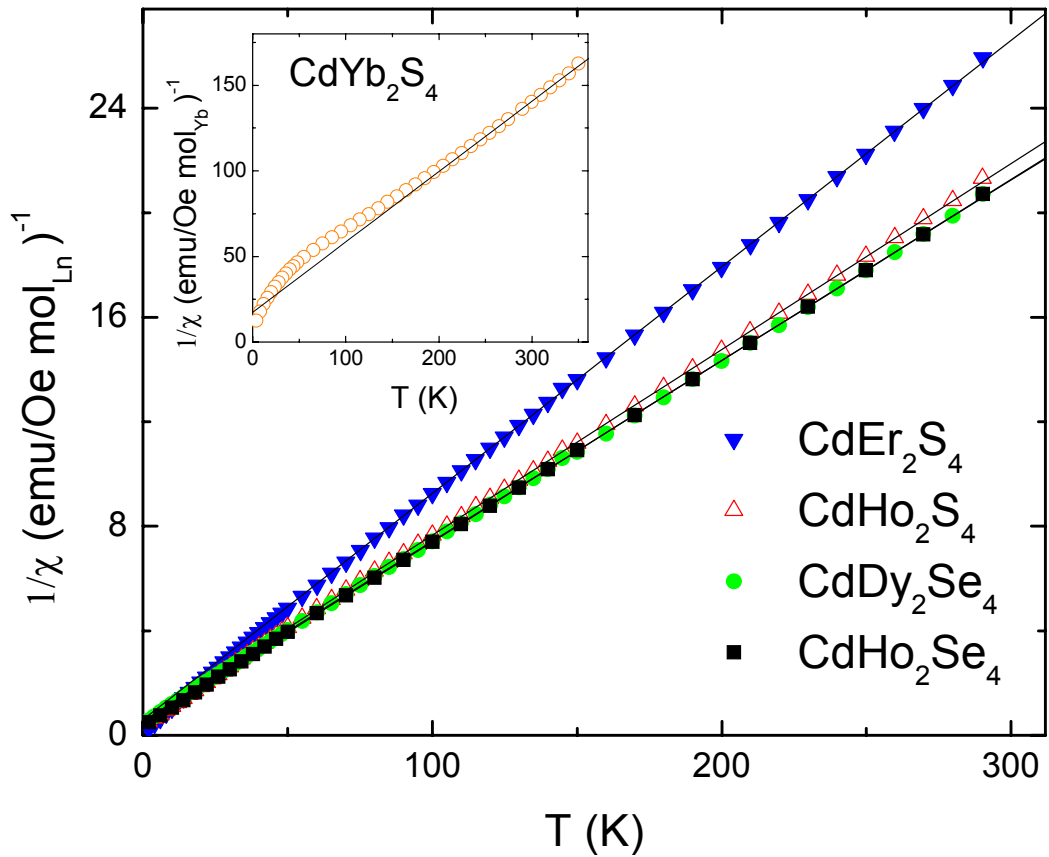

**Figure 3**.

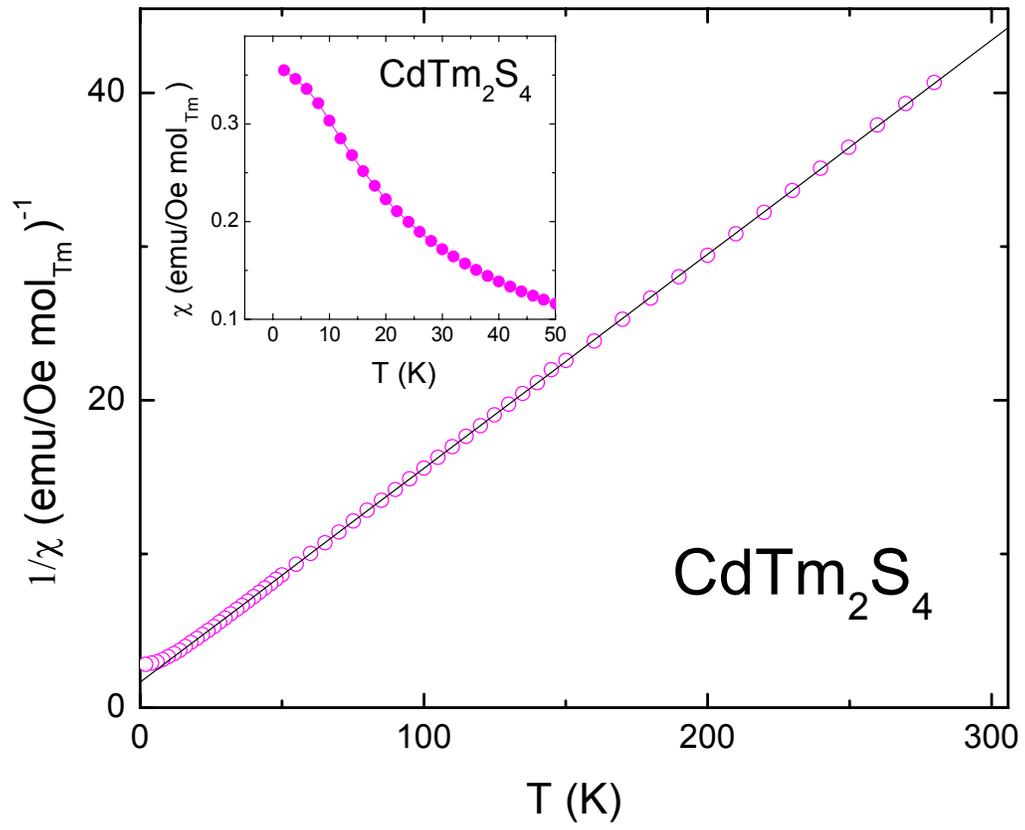

**Figure 4.**

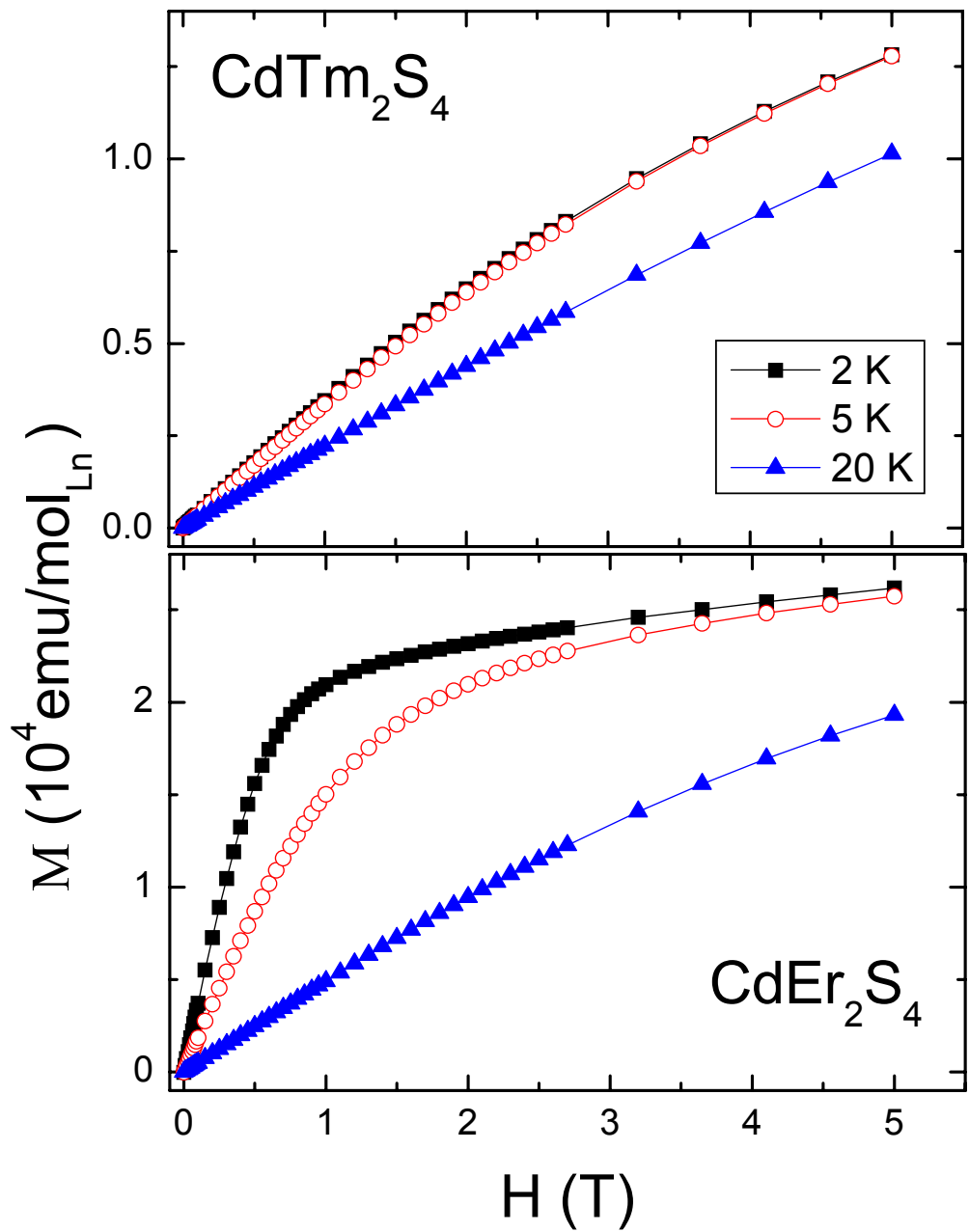

**Figure 5.**

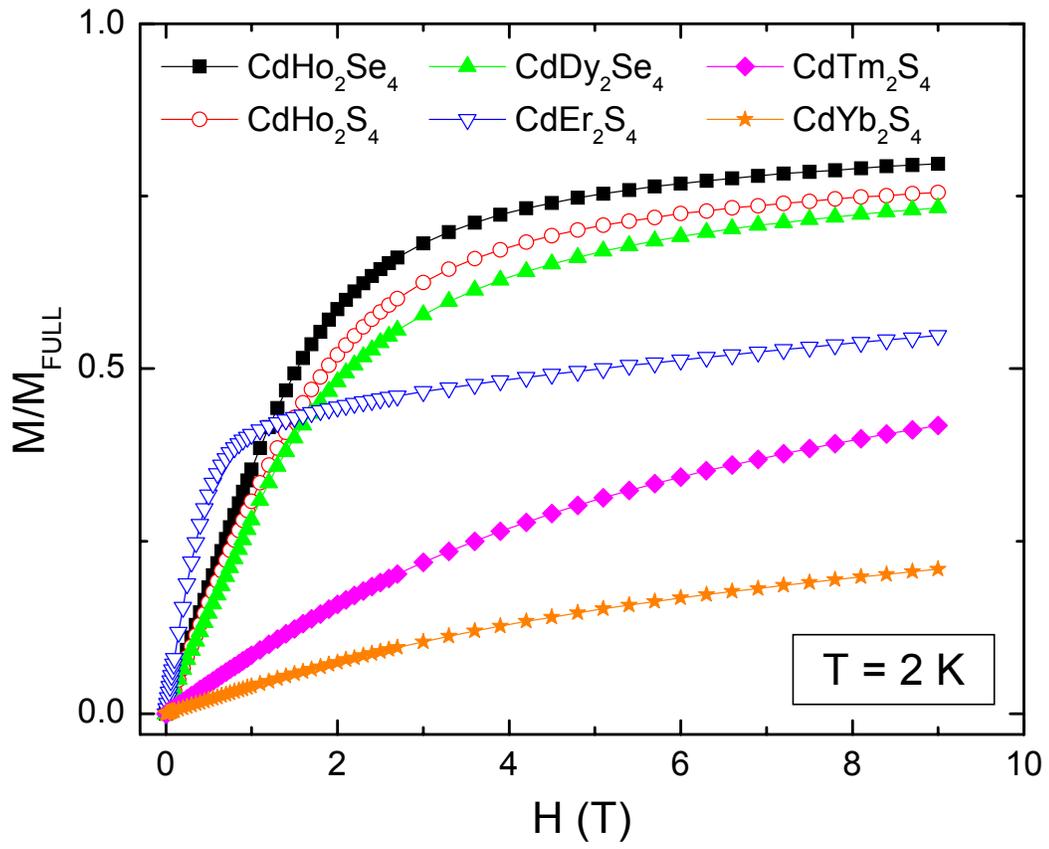

**References**

[1] A. P. Ramirez, in *Handbook of Magnetic Materials*, edited by K. J. H. Buschow (Elsevier, Amsterdam, 2001), Vol. 13.

[2] P. Schiffer and A. P. Ramirez, Comments Cond. Mat. Phys. **18**, 21 (1996).

[3] J. D. M. Champion *et al.*, Phys. Rev. B **68**, 020401(R) (2003).

[4] M. J. P. Gingras, C. V. Stager, N. P. Raju, B. D. Gaulin, and J. E. Greedan, Phys. Rev. Lett. **78**, 947 (1997).

[5] B. D. Gaulin, J. N. Reimers, T. E. Mason, J. E. Greedan, and Z. Tun, Phys. Rev. Lett. **69**, 3244 (1992).

[6] J. N. Reimers, J. E. Greedan, R. K. Kremer, E. Gmelin, and M. A. Subramanian, Phys. Rev. B **43**, 3387 (1991).

[7] J. S. Gardner *et al.*, Phys. Rev. Lett. **82**, 1012 (1999).

[8] J. S. Gardner *et al.*, Phys. Rev. B **68**, 180401(R) (2003).

[9] Y. Yasui *et al.*, J. Phys. Soc. Jpn. **71**, 599 (2002).

[10] S. T. Bramwell and M.J. Harris, J. Phys. Condens. Matt., **10**, L215 (1998).

[11] A. P. Ramirez *et al.*, Nature **399**, 333 (1999).

[12] J. Snyder, J. S. Slusky, R. J. Cava, and P. Schiffer, Nature **413**, 48 (2001); Phys. Rev. B **66**, 064432 (2002).

[13] J. W. Snyder *et al.*, Phys. Rev. B **69**, 064414 (2004).

[14] H. Fukazawa, R. G. Melko, R. Higashinaka, Y. Maeno, and M. J. P. Gingras, Phys. Rev. B **65**, 054410 (2002).

[15] T. Sakakibara, T. Tayama, Z. Hiroi, K. Matsuhira, and S. Takagi Phys. Rev. Lett. **90**, 207205 (2003).

[16] S. T. Bramwell, *et al.*, Phys. Rev. Lett. **87**, 047205 (2001).

[17] J. Snyder *et al.*, Phys. Rev. Lett. **91**, 107201 (2003); Phys. Rev. B **70**, 184431 (2004).

[18] S. T. Bramwell and M. J. P Gingras, Science **294**, 1495 (2001).

[19] R. Siddharthan *et al.*, Phys. Rev. Lett. **83**, 1854 (1999).

[20] R.G. Melko, B.C. den Hertog, and M. J. P. Gingras, Phys. Rev. Lett. **87**, 067203 (2001).

[21] S. T. Bramwell *et al.*, J. Phys.-Condens. Mat. **12**, 483 (2000).

[22] K. Matsuhira *et al.*, J. Phys.-Condens. Mat. **12**, L649 (2000).

[23] A. P. Ramirez *et al.*, Phys. Rev. Lett. **89**, 067202 (2002).

[24] J. A. Hodges *et al.*, Phys. Rev. Lett. **88**, 077204 (2002).

[25] J. S. Gardner *et al.*, Phys. Rev. B **64**, 224416 (2001).

[26] M. J. P. Gingras *et al.*, Phys. Rev. B **62**, 6496 (2000).

[27] H. Martinho *et al.*, Phys. Rev. B **64**, 024408 (2001).

[28] J. Ostorero *et al.*, Phys. Rev. B **40**, 391 (1989).

[29] L. Morellon *et al.*, Phys. Rev. B **58**, R14721 (1998).

[30] W. Choe *et al.*, Phys. Rev. Lett. **84**, 4617 (2000).

[31] O. M. Aliev, A. B. Agaev, and R. A. Azadaliev, Inorg. Mat. **33**, 1123 (1997).

[32] L. Bendor and I. Shilo, J. Solid State Chem. **35**, 278 (1980).

[33] J. Flahaut *et al.*, Acta Crystallogr. **19**, 14 (1965).

[34] A. Tomas *et al.*, Acta Crystallogr. Sect. B **42**, 364 (1986).

[35] A. Tomas, I. Shilo, and M. Guittard, Mat. Res. Bull. **13**, 857 (1978).

[36] A. Tomas *et al.*, Mat. Res. Bull. **20**, 1027 (1985).

[37] L. Bendor, I. Shilo, and I. Felner, J. Solid State Chem. **39**, 257 (1981).

[38] H. Fujii, T. Okamoto, and Kamigaic.T, J. Phys. Soc. Jpn. **32**, 1432 (1972).

[39] S. Kainuma, J. Phys. Soc. Jpn. **30**, 1205 (1971).

[40] S. Pokrzywnicki, Physica Status Solidi B-Basic Research **71**, K111 (1975).

[41] S. Pokrzywnicki and A. Czopnik, Phys. Status Solidi B **70**, K85 (1975).

[42] S. Pokrzywnicki, L. Pawlak, and A. Czopnik, Phys. B & C **86**, 1141 (1977).

[43] L. Pawlak, M. Duczmal, and A. Zygmunt, J. Magn. Magn. Mater. **76-7**, 199 (1988).

[44] O. Knop, F. Brisse, Castelli.L, et al., Can. J. Chem. **43**, 2812 (1965).